\documentclass[preprint]{aastex}

\usepackage{rotating}

\shorttitle{Inner Galaxy delineated by 6.7--GHz methanol masers}
\shortauthors{Green et al.}

\begin{document}

\title{Major structures of the inner Galaxy delineated by 6.7--GHz methanol masers}
\author{J. A. Green$^{1}$, J. L. Caswell$^{1}$, N. M. McClure-Griffiths$^{1}$,\\ A. Avison,$^{2}$ S. L. Breen,$^{1}$ M. G. Burton,$^{3}$
S. P. Ellingsen,$^4$ G. A. Fuller,$^2$\\ M. D. Gray,$^{2}$ M. Pestalozzi,$^{5}$ M. A. Thompson,$^6$ M. A. Voronkov$^1$}
\affil{$^{1}$ CSIRO Astronomy \& Space-Science, Australia Telescope National Facility, PO Box 76, Epping, NSW 1710, Australia}
\affil{$^{2}$ Jodrell Bank Centre for Astrophysics, Alan Turing Building, University of Manchester, Manchester, M13 9PL, UK}
\affil{$^{3}$ School of Physics, University of New South Wales, Sydney, NSW 2052, Australia}
\affil{$^{4}$ School of Mathematics and Physics, University of Tasmania, Private Bag 37, Hobart, TAS 7001, Australia}
\affil{$^{5}$INAF/IFSI, via del Fosso del Cabaliere 100, I-00133, Roma, Italy}
\affil{$^{6}$Centre for Astrophysics Research, Science and Technology Research Institute, University of Hertfordshire, College Lane, Hatfield, AL10 9AB, UK}

\begin{abstract}
We explore the longitude-velocity distribution of 6.7--GHz methanol masers in the context of the inner structure of our Galaxy. We analyse the correlation in velocities within this distribution and identify density enhancements indicating large-scale regions of enhanced star formation. These are interpreted as the starting points of the spiral arms and the interaction of the Galactic bar with the 3--kpc arms. The methanol masers support the presence of a long thin bar with a 45$^{\circ}$ orientation. Signatures of the full 3--kpc arm structure are  seen, including a prominent tangent at approximately $-$22$^{\circ}$ Galactic longitude. We compare this distribution with existing models of the gas dynamics of our Galaxy.  The 3--kpc arm structure appears likely to correspond to the radius of corotation resonance of the bar, with the bar on its inner surface and the starting points of the spiral arms on its outer surface. 
\end{abstract}

\keywords{masers --- physical data and processes, stars: formation --- stars, 
Galaxy: structure --- The Galaxy}

\section{Introduction}
Studying the structure and dynamics of the Milky Way is complicated by our location within it, and the resultant obscuration along the line-of-sight, but nevertheless it has established that our Galaxy has a grand spiral structure, with spiral arms, interior structures such as the 3--kpc arms \citep{vanwoerden57, dame08} and one or two Galactic bars (\citealt{blitz91}; \citealt{gerhard02}, references therein; \citealt{benjamin05}). Early descriptions of the structure of our Galaxy were inferred from star counts and positioning of objects with photometric distances, but current revisions have been largely based upon the kinematics of atomic and molecular tracers, such as H{\sc i} and CO. The sinusoidal distribution of velocities with Galactic longitudes has established the approximately circular and symmetric rotation of the Galaxy and both tracers have been used to constrain dynamical models \citep[e.g.][]{peters75,burton78b,liszt78, binney91,weiner99, fux99}. However, there are limitations to this approach. H{\sc i} is detected throughout the Galaxy in such great abundance that only the highest or lowest velocities along a given line of sight (the terminal velocities and spiral arm tangent deviations) can be used to constrain dynamical modelling. CO is less diffuse and more concentrated in the major structural features of our Galaxy, but surveys of CO are limited by a lack of sensitivity to material on the far side of the Galaxy. 
An alternative tracer, the population of 6.7--GHz methanol masers, has the advantage of only being observed towards regions of high-mass star formation \citep{pestalozzi02b,minier03,sridharan02,walsh03,xu08} and the strong and narrow emission of these masers is well correlated with their systemic velocity, the central maser velocity typically within 3--5\,km\,s$^{-1}$ \citep{szymczak07, pandian09}. They are also found in abundance throughout the Galaxy \citep{pestalozzi05} and have been detected to Galactocentric distances of 13.5\,kpc \citep{honma07}. Inner Galaxy structures have been shown to be traced by 6.7--GHz methanol masers: 45 sources are associated with the 3--kpc arms \citep{green10mmb2}; 7 sources with the high velocities associated with Galactic bars \citep{caswell97, caswell10mmb1,green10mmb2}; and 11 with Sagittarius B2 \citep[][references therein]{caswell10mmb1}. The 6.7--GHz methanol masers are excellent measures of the kinematic behaviour of the structure of our Galaxy within the terminal velocities outlined in H{\sc i} and CO. 

In this paper we explore how 6.7--GHz methanol masers trace the major star forming structures of the inner Galaxy, specifically: the Galactic bar, the 3 kpc arms, and the origins of the spiral arms. All of these structures are believed to be confined within $\sim$4 kpc of the Galactic centre. Assuming a solar distance of 8.4 kpc \citep{ghez08}, the inner 4 kpc of the Galaxy is geometrically contained within the Galactic longitude range $-$28$^{\circ}$ to $+$28$^{\circ}$. The Methanol Multibeam (MMB) survey \citep{green09a}, which observed from $-$174$^{\circ}$ to $+$60$^{\circ}$ with the Parkes Radio Telescope, detected $\sim$550 masers within longitudes $\pm$28$^{\circ}$. The MMB had a 3 $\sigma$ sensitivity of 0.7 Jy, covered the latitude range of $\pm$2$^{\circ}$ and had a velocity coverage encompassing the extent of CO emission seen in \citet{dame01}. The MMB catalogue therefore represents the most complete survey of current high-mass star formation traced by 6.7-GHz methanol masers in the region, providing a census of the regions which includes their kinematic behaviour. 
Through examining  the distribution and densities of these masers in longitude and velocity, in comparison with the spiral arms, we attempt to delineate the inner Galactic structure. 

\section{6.7--GHz methanol maser distribution and density enhancements}\label{lvintro}
\begin{figure}
 \begin{center}
 \renewcommand{\baselinestretch}{1.1}
\includegraphics[width=16cm]{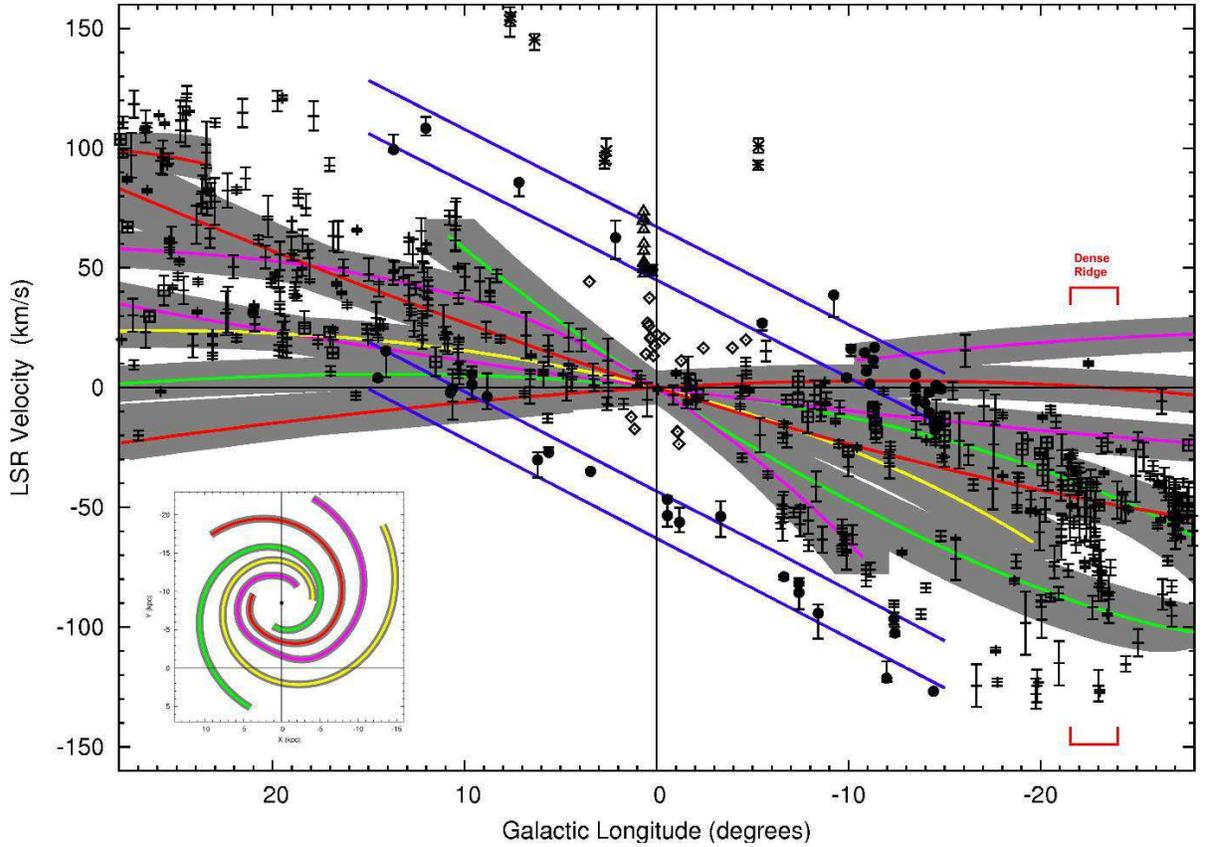} 
\caption{\small Longitude-velocity `crayon' plot showing the distribution of 6.7--GHz methanol masers within $\pm$28$^{\circ}$ overlaid with example spiral arm loci (the spatial pattern of \citet{taylor93}, transferred to the {\it l,v} domain by the \citet{brand93} rotation curve). Coloured loci are the spiral arms defined by the model, grey shading incorporates an arm thickness of 1 kpc and a velocity tolerance of 7\,km\,s$^{-1}$. Yellow loci represent the Perseus spiral arm; Purple - Carina-Sagittarius; Orange - Crux-Scutum; Green - Norma. The blue lines delineate the region identified in CO emission as the 3--kpc arms by \citet{dame08}. Crosses show 6.7--GHz methanol masers of the MMB survey. Circles are masers associated with the 3--kpc arms \citep[see][for details]{green09b,caswell10mmb1,green10mmb2}. Diamonds are masers which are interior to the 3--kpc arms, primarily candidates for belonging to the Galactic Centre Zone \citep{caswell10mmb1}. Triangles are masers associated with the Sagittarius B2 complex. Stars are masers associated with the Galactic bar. Crosses enclosed in squares are masers with high latitudes (and therefore likely to be closer to us than 4.5\,kpc, see Section \ref{lvintro}). The starting points of the spiral arms have been adjusted from \citet{taylor93} to match the discussion of Section \ref{armorigins}. The red brackets highlight the longitude range of the dense ridge of sources discussed in the text.}
\label{lvdistribution}
\end{center}
\end{figure}

\begin{figure}
 \begin{center}
 \renewcommand{\baselinestretch}{1.1}
\includegraphics[width=13cm]{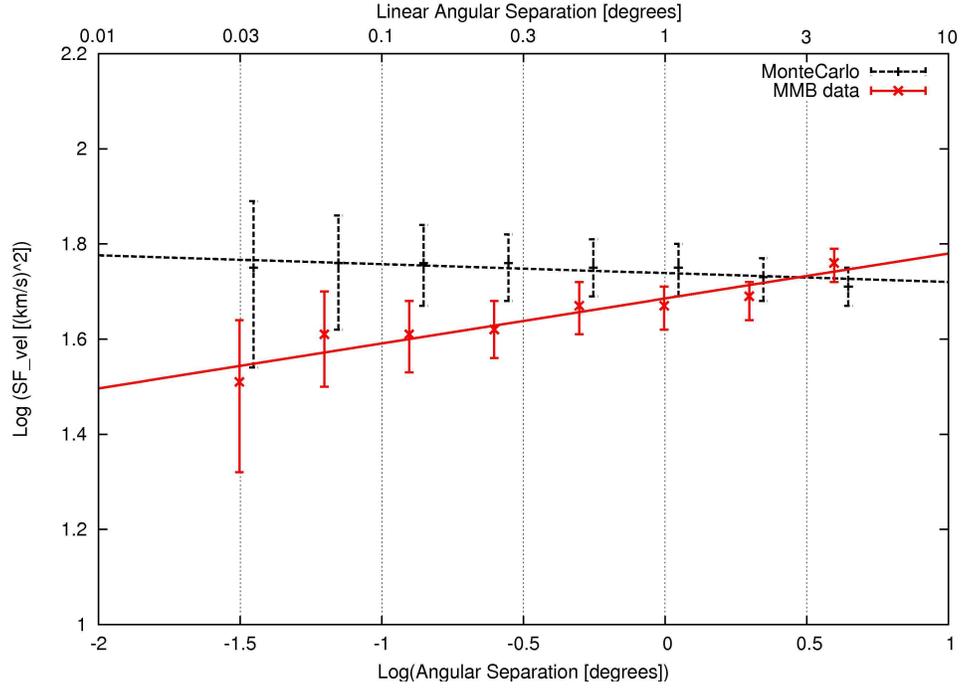}
\caption{\small Velocity structure function showing the correlation in velocity for given angular separations of maser sites. The data is binned in equal logarithmic bins (0.3) and compares all the MMB data between $\pm$28$^{\circ}$ longitude (red) with a Monte Carlo simulation (black). The Monte Carlo simulation has been systematically offset in angular separation by +0.05 for clarity of comparison. The error bars represent the statistical (Poisson) error in the mean of each bin.}
\label{StrutFunkfigure}
\end{center}
\end{figure}

\begin{figure}
 \begin{center}
 \renewcommand{\baselinestretch}{1.1}
\includegraphics[width=15cm]{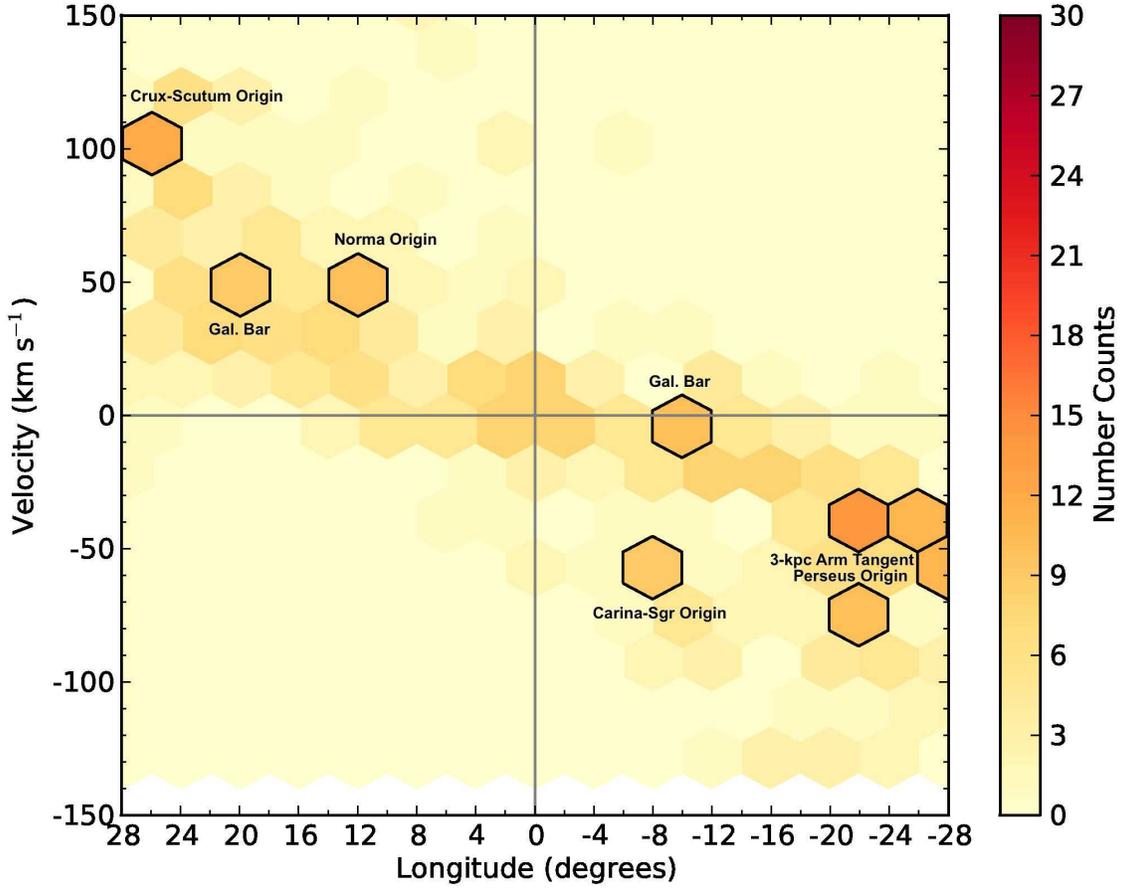}
\caption{\small Longitude-velocity `blockbusters' plot showing the density of masers within the {\it l,v} domain. As described in the text the distribution of sources seen in Figure \ref{lvdistribution} has been smoothed on scales $<$0.03$^{\circ}$ and then binned on scales $\ge$3$^{\circ}$. Hexagonal bins were used as they reduce the visual distortion in the distribution caused by binning. The colour-scale represents the number of sources. Black lines delineate statistically significant dense regions, at the 5\,$\sigma_{poisson}$ level. Labels correspond to enhanced star formation at the ends of the Galactic bar and the starts of the spiral arms.}
\label{lvdensity}
\end{center}
\end{figure}

\begin{figure}
 \begin{center}
 \renewcommand{\baselinestretch}{1.1}
\includegraphics[width=15cm]{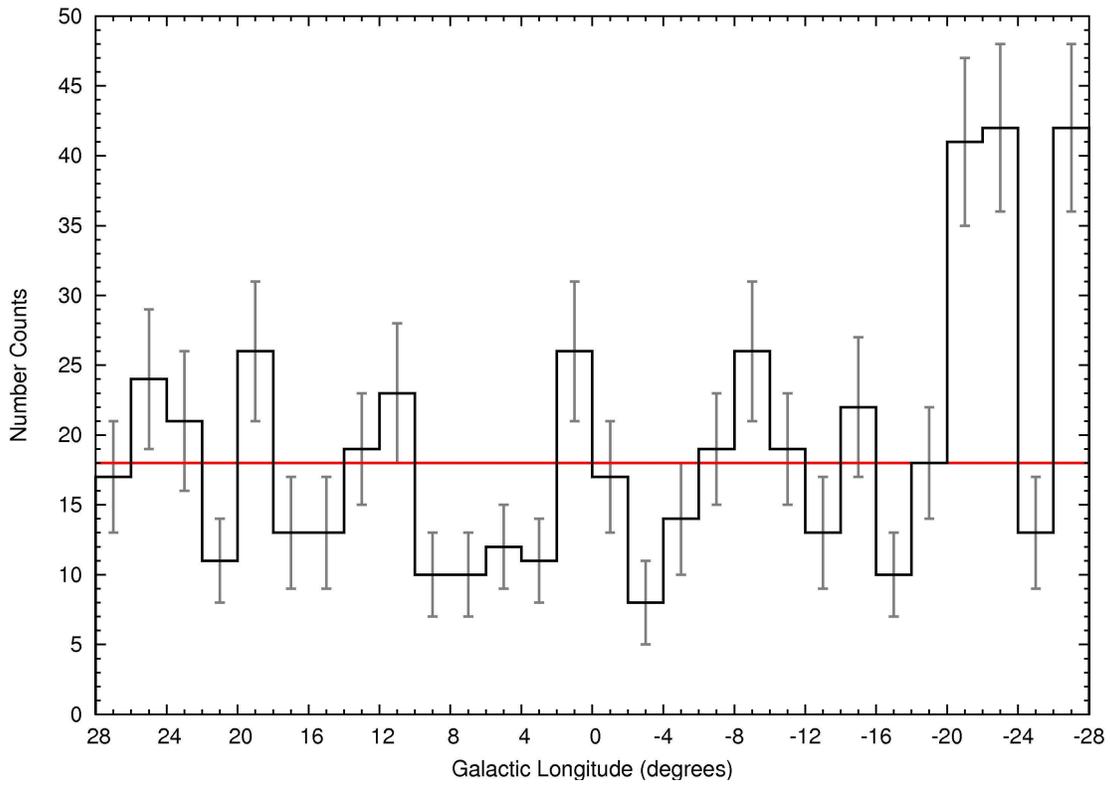}
\caption{\small Distribution of sources with longitude in 2$^{\circ}$ bins. Median number count per bin is 18 (red line). Errors represent Poisson statistical error margins. Beyond the limits of the figure, existing maser data suggest the number counts decline to (or below) the average beyond $-$28$^{\circ}$, but show a peak at positive longitudes, beyond +28$^{\circ}$.}
\label{LCOUNTS}
\end{center}
\end{figure}

The longitude-velocity ({\it l,v}) distribution of 6.7--GHz methanol masers between longitudes $\pm$28$^{\circ}$ is shown in Figure\,\ref{lvdistribution}. It is immediately apparent by eye that the maser distribution is not entirely random, but contains clumps of sources on the sub-degree scale over a few kilometres per second, together with structures on the scale of several degrees over tens of kilometres per second. The former is due to multiple sources within molecular cloud complexes, whilst the latter is a signature of Galactic scale structures. A structure function of the {\it l,v} distribution formally demonstrates the presence of this structure, quantifying the velocity correlation for given angular separations. We define the second order structure function of velocity as a function of angular separation:

\begin{equation}
\label{funkeq}
SF_{vel}(r)=\langle[vel(x)-vel(x+r)]^2\rangle
\end{equation}

\noindent
where {\it x} is the angular position of a maser site and {\it r} is the angular separation between a pair of maser sites. We take the ensemble average of measurements with the same range of angular separation (denoted by the angular brackets). The resulting structure function is shown in Figure \ref{StrutFunkfigure} for both the masers and a Monte Carlo simulation together with statistical errors. 
The Monte Carlo simulation is a random distribution of masers in the plane of the Galaxy with longitudes between $\pm$28$^{\circ}$ and Galactocentric distances between 3 and 13.5 kpc. The masers were assigned velocities based on a Galactocentric solar distance of 8.4 kpc \citep{ghez08,gillessen09} and flat rotation curve with a circular rotation of the Sun of 246\,km\,s$^{-1}$ \citep{reid09,bovy09}. 
We generated 482 sources by this method, to match the observed quantity for the {\it l,v} domain, i.e. not including the 63 masers mentioned in the introduction associated with the 3 kpc arms (45), Galactic bar (7), and Sagittarius B2 (11). A power law fit to the structure function gives a gradient in the log-log plane of $+$0.09$\pm$0.01 for the masers and $-$0.02$\pm$0.01 for the Monte Carlo simulation (with the errors the formal errors in the fit). The consistent positive gradient of the structure function of the data is indicative of a clumpiness of structure on large scales (both angular and in velocity)  and a correlation on small scales. In comparison, the approximately flat structure function of the Monte Carlo simulation demonstrates the same level of velocity correlation at all scales, i.e. no scale-specific structure. The measured structure function departs from the simulation, within the errors (both formal and statistical), at scales less than 0.03$^{\circ}$ (with correlated velocities up to 30\,km\,s$^{-1}$) and at scales greater than 3$^{\circ}$ (with correlated velocities up to 50\,km\,s$^{-1}$). Smoothing the distribution of sources on the small scales (by counting all sources within 0.03$^{\circ}$ and 30\,km\,s$^{-1}$ singularly) and binning the data on the large scales we produce the {\it l,v} density distribution seen in Figure \ref{lvdensity}. The non-zero mean of the bins is 3.7 sources, giving a Poisson noise of 1.9 sources. At a 5$\sigma$ confidence level, statistically significant density fluctuations are therefore those with 9.7 or more sources, and these are highlighted in Figure \ref{lvdensity}. The density enhancements are also apparent in the distribution of number counts with longitude, shown in Figure \ref{LCOUNTS}.

The majority of the Galactic scale structure seen in the 6.7--GHz methanol maser {\it l,v} domain is interpretable as the spiral arms. Figure\,\ref{lvdistribution} overlays the masers on the loci of the spiral arms commonly adopted in analyses of structure in the {\it l,v} domain: the four arm shape of \citet{taylor93}, effectively the same as that of \citet{georgelin76}, transferred to the {\it l,v} domain using the rotation curve of \citet{brand93}. We adopt the solar rotation and radius as mentioned in the previous paragraph and use these throughout the discussion. The choice of rotation curve, solar parameters and spiral arm shape have minimal effect, but these are addressed in detail in the appendix. Sources already associated with star forming regions towards the Galactic centre (Sgr B2, the Galactic Centre Zone and sites within 3.5--kpc of the Galactic centre, see \citet{caswell10mmb1} for details) and the parallel sections of the 3--kpc arms between longitudes $\pm$15$^{\circ}$ are distinguished by different symbols. With these excluded from consideration, then incorporating an arm thickness of 1 kpc with a velocity tolerance of $\pm$7\,km\,s$^{-1}$, we find the spiral arms account for 79\% of the masers. The Monte Carlo simulation (from 100 realisations) gives an average (both mean and median) number of associated sources of 69\% with a standard deviation of 2\%, implying the association of real masers is statistically significant (none of the Monte Carlo simulations exceed 72\% association). We chose an arm thickness of 1 kpc based on estimates of the inter-arm separation and typical widths of the arms in models \citep[see for example][]{gomez04, sewilo04,mcclure04,levine06}. Varying the arm thickness between 0.5 and 1.5 kpc results in the level of association varying between 49 and 78\%. Our velocity tolerance of $\pm$7\,km\,s$^{-1}$ is chosen based on the kinematic uncertainty of the masers in relation to the high-mass star forming region they are tracing  \citep{reid09, pandian09}. Varying the velocity tolerance between 5\,km\,s$^{-1}$ and 10\,km\,s$^{-1}$ results in an association of between 58 and 80\%. The unassociated sources (approximately 20\%) lie almost exclusively in the longitude regions 18$^{\circ}$ to 28$^{\circ}$ with velocities $>$100\,km\,s$^{-1}$ and between $-$19$^{\circ}$ and $-$24$^{\circ}$ with velocities between 0 and $-$130\,km\,s$^{-1}$. These sources will be discussed in the following sections.

\begin{figure}
 \begin{center}
 \renewcommand{\baselinestretch}{1.1}
\includegraphics[width=9cm]{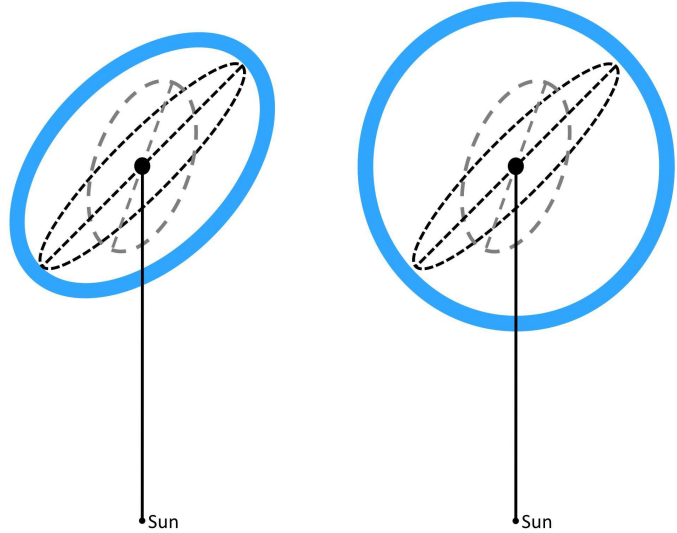} 
\caption{\small Schematic of the two simple interpretations of the 3--kpc arms. The blue represents the 3-kpc arms as either an elliptical structure or a ring. The black line is the Sun-Galactic Centre. Black short dashes represent the long bar and grey long dashes represent the short bar.}
\label{barschematic}
\end{center}
\end{figure}

\begin{table*} 
\centering 
\caption{ Parameters of density enhancements of 6.7--GHz methanol masers in the longitude-velocity domain. For explanation of source density see main text. Densities with $^\dagger$ are the sum of more than one enhancement, confused by multiple passes of spiral arms along the line of sight. The last column lists Galactic structures which are believed to be responsible for the enhancement.} 
\begin{tabular}{cccccccl}
\hline
\multicolumn{2}{c}{Longitude Range} & \multicolumn{3}{c}{Latitude Range} & \multicolumn{1}{c}{Velocity} & \multicolumn{1}{c}{Source} & \multicolumn{1}{l}{Associated}\\
{\it l} min & {\it l} max & {\it b} min & {\it b} max & {\it b} med & Range & Density & Galactic Structures\\
( $^{\circ}$) & ( $^{\circ}$) & ( $^{\circ}$) & ( $^{\circ}$) & ( $^{\circ}$) & ( \,km\,s$^{-1}$) & (kpc$^{-2}$) & \\
\hline
+24 & +28 & $-$0.3 & +0.3 & 0.1& $>$90 & 11 & Crux-Scutum arm origin\\
+18 & +22 & $-$0.5 & +0.1 & $-$0.1&+40,+65&15& near-side Galactic Bar?\\
+10 & +14 & $-$0.6 & +0.5 & $-$0.1& 0,+70 &38$^\dagger$&  Norma arm origin?\\
&&&&&&&all spiral arms\\
$-$4 & +6 & $-$0.5 & $+$0.3 & $-$0.1& $-$20,+20 &18$^\dagger$& Sgr B2, GCZ, \\
&&&&&&&all spiral arms\\
&&&&&&&all solar circle masers\\
$-$11 & $-$6 & $-$0.2 & +0.2 & 0.0 & $-$45,$-$70 &15& Carina-Sgr arm origin\\
$-$15 & $-$9 & $-$0.5 & +0.4 & 0.0 & $-$30,+10 &16& Far 3--kpc arm \\
&&&&&&&\& far-side Galactic bar \\
$-$24 & $-$20& $-$0.5 &+0.6 & $-$0.1 & $-$30,$-$95&48$^\dagger$& 3--kpc arm tangent, \\
&&&&&&&Perseus arm origin, \\
&&&&&&&\& Norma, Crux-Scutum arms \\
$-$28 & $-$24 & $-$0.7 & +0.8& $-$0.2 & $-$35,$-$70 &20$^\dagger$& Crux-Scutum \\
&&&&&&&\& Norma arms \\
\hline
\end{tabular} 
\label{resotable}
\end{table*}

\section{Structural interpretation of {\it l,v} distribution}
The {\it l,v} distribution raises the question: what are the causes of the statistically significant over-densities $-$ are they related to Galactic structure or random fluctuations in star formation? Regions of star formation are contained within Giant Molecular Clouds (GMCs) which have physical scales of 10s of parsecs. With the high-mass star formation traced by the 6.7--GHz masers, we are looking at regions at typical distances of $\ge$2 kpc (the {\it l,v} distribution within $\pm$28$^{\circ}$ suggesting mostly $\ge$4 kpc). At these distances most GMCs subtend angular scales smaller than 1$^{\circ}$ . The coherent dense regions of sources seen in Figure \ref{lvdistribution}, and listed in Table \ref{resotable}, are on scales of $>$1$^{\circ}$ (highlighted by their presence post-smoothing in Figure \ref{lvdensity}). This makes random fluctuations, unrelated to Galactic scale processes, an unlikely cause of the increased star formation. This section analyses the 6.7--GHz methanol maser {\it l,v} distribution in relation to the major structural features of the inner Galaxy. We begin with the Galactic bar, then the 3--kpc arms, and follow with the starting points of the spiral arms, or the spiral arm `origins'. Excluded from the discussion are masers at high latitudes, as these are unlikely to lie within 4\,kpc of the Galactic centre. With a scale height of star formation estimated to be $\sim$30$-$40\,pc \citep{fish03}, any masers whose latitudes imply a Galactic height, z, greater than $\pm$60$-$80\,pc at a heliocentric distance of 4.4\,kpc can be excluded from inner Galaxy considerations. This corresponds to latitudes 0.8$^{\circ}$ to 1.0$^{\circ}$ and we therefore adopt 0.9$^{\circ}$ as a cut-off. Within the longitude region $\pm$28$^{\circ}$ there are 29 masers with latitudes in excess of 0.9$^{\circ}$ from the plane.  

 \subsection{The Galactic bar}\label{barsection}
The Galactic bar has usually been regarded as either a triaxial boxy bulge (or `peanut'), principally on the basis of Cosmic Background Explorer (COBE) infrared observations, or as a long thin bar. \citet{blitz91} speculated that the Galaxy might contain two bar components and recent consensus is that both these types of bars exist \citep{hammersley00, benjamin05,cabrera08}. The orientation of the bar with respect to the Sun-Galactic centre line-of-sight is thought to lie between 14$^{\circ}$ and 45$^{\circ}$, but many of the lower inclination angles are for boxy bulge bar models, whilst the larger angles are for long thin bars (and confusion has been added with the interchangeable and sometimes ambiguous use of the bulge and bar terms, see discussions of \citet{ng98} and \citet{lopez99}). This has led to a likely picture of a triaxial bulge bar inclined at 20$-$30$^{\circ}$ and a long thin bar at $\sim$45$^{\circ}$. This combination is shown relative to the 3-kpc arms in Figure \ref{barschematic}. Primarily on the basis of models based on infrared (IR) data, the radius of corotation resonance of the bar (of the order of 1.2 times the semi-major axis of the bar \citep{elmegreen96, englmaier06, buta09}) is estimated to be 4$\pm$0.5\,kpc  \citep[][references therein]{gerhard02}. In galaxies this radius marks the transition from elliptical orbits following the bar to circular orbits beyond \citep{contopoulos80}. As noted in the introduction, several 6.7--GHz methanol masers have been associated with the Galactic bar. Where the Galactic bar and the 3--kpc arms meet (discussed in section \ref{3kpcdiscussion}), a region of enhanced star formation is expected \citep[e.g.][]{fux99, englmaier99, lopez01, binney08}. This means that a density enhancement in either the near or far 3--kpc arm 6.7--GHz methanol maser population could indicate the interaction with the end of the bar. Such an enhancement has been suggested in the far 3--kpc arm population \citep{green10mmb2} and we now address both this and possible near-side interactions.

\subsubsection{Density enhancement at the far end of the bar}
The 6.7--GHz methanol maser population tracing the 3--kpc arms exhibits a high density of sources within the far 3--kpc arm between longitudes $-$9$^{\circ}$ to $-$15$^{\circ}$ (see Figure \ref{lvdensity} and Table \ref{resotable}). Within this dense region, 8 sources with velocities $>$0\,km\,s$^{-1}$ partially overlap with the extrapolated outer Galaxy components of the Carina-Sagittarius and Crux-Scutum arms (at distances $>$15\,kpc), and 6 sources with velocities between 0 and $-$30\,km\,s$^{-1}$ overlap with the Carina-Sagittarius arm (which passes between the Galactic centre and the Sun at a distance of $<$1.1 kpc \cite{sato10}). 

The distances of these arms will have an impact on the flux density distribution of the sources, an effect that can be estimated if we assume a simple power law luminosity distribution \citep[e.g.][]{pestalozzi07} above the sensitivity limit of the MMB (0.7\,Jy, \citealt{green09a}): at 15\,kpc heliocentric distance only the brightest sources would be detected, whilst at 1 kpc heliocentric distance very faint sources would be. If the far 3--kpc arm sample of sources was significantly biased by distant sources, we would expect a shallow distribution of peak flux densities, whilst if it was biased by nearby sources, we would expect an asymmetric distribution with a peak at very low peak flux densities. We do not see either, instead finding an approximately Gaussian distribution with a median of 2.4 Jy, implying a population dominated by intermediate distances. Furthermore, the far spiral arms are not well traced by H{\sc i} or CO at these longitudes, so their location is less precise. Regions of high-mass star formation are comparatively rare, so without either of the spiral arms being tangential, we would not expect there to be a significant population present in either. These factors combine to give us  confidence that the dense region identified by \citet{caswell10mmb1} and \citet{green10mmb2} is a product of far 3--kpc arm sources interacting with the bar.

\subsubsection{Density enhancement at the near end of the bar: implications for length and orientation}\label{barimp}
Expanding on the analysis of \citet{green10mmb2}, we assume a bar orientation of 45$^{\circ}$ \citep[e.g.][]{benjamin05}, with the semi-major axis left as a free parameter. The density enhancement at the far end of the bar (between longitudes $-$9$^{\circ}$ and $-$15$^{\circ}$) thus implies a semi-major axis of the bar between 2.2 and 4.3 kpc. This indicates that the near end of the bar lies between longitudes 13$^{\circ}$ and 30$^{\circ}$. We see from Table \ref{resotable} and Figure \ref{lvdensity} that there are three higher density regions of masers within this range: between 10$^{\circ}$ and 14$^{\circ}$ (1 bin above 5\,$\sigma_{poisson}$); between 18$^{\circ}$ and 22$^{\circ}$ longitude (1 bin above 5\,$\sigma_{poisson}$); and between 24$^{\circ}$ and 28$^{\circ}$ longitude (1 bin above 5\,$\sigma_{poisson}$); these correspond to semi-major axes of 1.8--2.4\,kpc, 2.9--3.4\,kpc and 3.7--4.1\,kpc respectively. 
In dynamical simulations, bars with short semi-major axes (or the assumption of such a mass distribution) are almost exclusively associated with small bar orientation angles \citep[e.g.][]{binney91,freudenreich98, babusiaux05} and a short bar with an acute orientation angle would produce densities in the far 3--kpc arm at longitudes smaller than is observed \citep{green10mmb2}. Additionally, a shorter bar of this type is associated with the Galactic bulge, but the 6.7--GHz methanol maser population has a narrow latitude distribution unassociated with the bulge \citep{caswell10mmb1}. Hence the semi-major axis is unlikely to be in the range 1.8--2.4 kpc. The estimate of a radius of corotation resonance at $\sim$4\,kpc implies that a semi-major axis of 3.7--4.1\,kpc is also unlikely. 

If, on the other hand, we assume the semi-major axis of the bar is fixed at 3.5\,kpc (the estimate of \citealt{gerhard02} and references therein, scaled to 8.4 kpc) and the bar orientation is instead left as a free parameter, the density enhancement at the far end of the bar implies a bar orientation between 35$^{\circ}$ and 53$^{\circ}$. This range of orientation angles would locate the near end of the bar between longitudes 20$^{\circ}$ and 24$^{\circ}$, which overlaps with the high density of masers seen at velocities close to +50\,km\,s$^{-1}$ (1 bin above 5\,$\sigma_{poisson}$ in Figure \ref{lvdensity}). 

In summary we believe the maser population traces the influence of a long thin bar with a semi-major axis of $\sim$3.4 kpc and an orientation of $\sim$45$^{\circ}$. This implies that, if both a long and short bar exist within our Galaxy, it is the long component which is primarily traceable by (high-mass) star formation and is younger. The short boxy/bulge bar was primarily identified in the IR observations of evolved stars and lacks 6.7--GHz methanol maser emission, thereby indicating it traces an older, more evolved population. Hence, our observations agree with the implied age dichotomy in the bars \citep[e.g.][]{lopez01}.

\subsection{The 3--kpc arms}\label{3kpcdiscussion} 
The (near) 3--kpc arm was originally discovered by \citet{vanwoerden57} as an absorption feature with a radial velocity of approximately $-$50\,km\,s$^{-1}$ at 0$^{\circ}$ longitude. It was named by \citet{oort58} based on the perceived H{\sc i} tangent point at longitude $-$22$^{\circ}$ corresponding to a Galactocentric radius of 3\,kpc (when the solar distance is 8.2\,kpc). At this longitude the distinct feature blends with spiral arm emission. Subsequently the negative longitude tangent has also been inferred as the cause for peaks in the longitudinal brightness profile of radio continuum at low frequencies (e.g. \citet{beuermann85} analysis of \citet{haslam81,haslam82} 408 MHz data), 2.4 $\mu$m emission (e.g. \citealt{hayakawa81}), OH/IR star kinematics \citep[e.g.][]{sevenster99b} and IR star counts \citep[e.g.][]{churchwell09}. However the work of \citet{hayakawa81} showed that the peak was not evident in higher frequency radio continuum emission, such as at 1.4 GHz \cite[e.g.][]{mathewson62} and 5 GHz \cite[e.g.][]{lockman79}. A positive longitude tangent has never been clearly observed. \citet{bania80} inferred a tangent at 23.6$^{\circ}$ (with a velocity of $\sim$110\,km\,s$^{-1}$) by assuming a ring-like structure, based on the ring fit of \citet{cohen76} for the H{\sc i} data across the longitude range 355$^{\circ}$ to 6$^{\circ}$. They found CO emission at $\le$110\,km\,s$^{-1}$, but it was not distinct from the rest of the Galactic emission. The question remains as to why we do not see a clear positive longitude tangent. One possible explanation is that it could be due to greater obscuration along the line of sight, with a larger length of the Carina-Sagittarius and Crux-Scutum arms present, compared with the negative tangent (where the spiral arms have an orientation almost perpendicular to the line of sight).

Recently \citet{dame08} discovered the long speculated far 3--kpc arm counterpart in CO emission, tracing both arms between $\pm$15$^{\circ}$ longitude. 
Forty-five 6.7--GHz methanol masers were associated with these parallel {\it l,v} structures \citep{green09b, caswell10mmb1} and one of these, the brightest known methanol maser 9.621+0.196, has a distance of 5.2$\pm$0.6\,kpc determined by astrometric parallax \citep{sanna09}. This places the maser at a Galactocentric distance of 3.4\,kpc, concurring with our expectations (see previous section) for the Galactocentric radius of the 3--kpc arms. The astrometric observations also showed the source to have a velocity component in a radial direction from the Galactic centre (41\,\,km\,s$^{-1}$) and a velocity component counter to the authors chosen model of Galactic rotation ($-$60\,km\,s$^{-1}$ relative to a flat rotation curve with circular velocity of 254\,km\,s$^{-1}$).

\subsubsection{A 3--kpc arm tangent}\label{3kpctan}
The methanol maser distribution demonstrates a large region with a high density of sources between $l = $ $-$18$^{\circ}$ to $-$28$^{\circ}$ (Figure \ref{lvdensity}). This region comprises a dense ridge (in velocity) of sources near $l = $ $-$22$^{\circ}$ and a further clump of sources (with a smaller range of velocities) near $l = $ $-$27$^{\circ}$ (see Figures \ref{lvdistribution} and \ref{LCOUNTS}). The clump of sources around $l = $ $-$27$^{\circ}$ is dominated by the G333 giant molecular cloud and is coincident with both the Norma arm (close to its tangent) and the Crux-Scutum arm and thus is excluded from the present analysis. The dense ridge near $l = $ $-$22$^{\circ}$ however, extends over almost 80\,km\,s$^{-1}$ of velocity, most of which does not align with common logarithmic spiral arm model loci. The Crux-Scutum arm lies at velocities less negative than $-$50\,km\,s$^{-1}$ whilst the Norma arm lies at velocities more negative than $-$100\,km\,s$^{-1}$ and less negative than $-$50\,km\,s$^{-1}$, leaving masers in the velocity range of $-$60 to $-$85\,km\,s$^{-1}$ unaccounted for. The only arm in conventional spiral arm models which could be extended into this region is the Perseus arm (see next section), but by itself this cannot account for the high density and range of velocities observed. The implication is hence that the ridge feature is a signature of the 3 kpc arm tangent.  The prominent ridge in the maser distribution also has (less obvious) counterparts in the {\it l,v} distributions of CO and H{\sc i} emission (as originally suggested by \citet{oort58}), H{\sc ii} regions \citep[e.g.][]{georgelin76, downes80, caswell87a}, pre-ultra compact H{\sc ii} region massive young stellar objects \citep[e.g.][]{urquhart07}  and ammonia emission (\citealt{walsh08} and A. Walsh, private communication). 

\subsubsection{Fitting an overall structure}\label{3kpcstrutfit}
Combining the density enhancement of masers near longitude $-$22$^{\circ}$ with the masers already assigned to the 3--kpc arms we have a distribution of two parallel structures between $\pm$15$^{\circ}$, which join at longitudes between $-$15$^{\circ}$ and $-$25$^{\circ}$,  over a range of $\sim$100\,km\,s$^{-1}$. Although the ridge is a strong signature for a tangent at negative longitudes, the positive longitude tangent is less obvious.  There are three regions of enhanced densities at positive longitudes (at approximately +12$^{\circ}$, +20$^{\circ}$ and +26$^{\circ}$, Figure \ref{lvdensity}), but they are all confused by spiral arm loci: the region at +12$^{\circ}$ is coincident with loci of every spiral arm; the region at +20$^{\circ}$ is overlapped by the loci of the Perseus, Carina-Sagittarius and Crux-Scutum arms; and the region at +26$^{\circ}$ is largely coincident with the Crux-Scutum arm (and will be discussed in the next section). The region at +20$^{\circ}$, which may be influenced by the Galactic bar (Section \ref{barsection}), includes a couple of sources outside the spiral arm parameter space and potentially extends to a broad velocity range of 15\,km\,s$^{-1}$ to 80\,km\,s$^{-1}$.

\begin{figure}
 \begin{center}
 \renewcommand{\baselinestretch}{1.1}
\includegraphics[width=12cm]{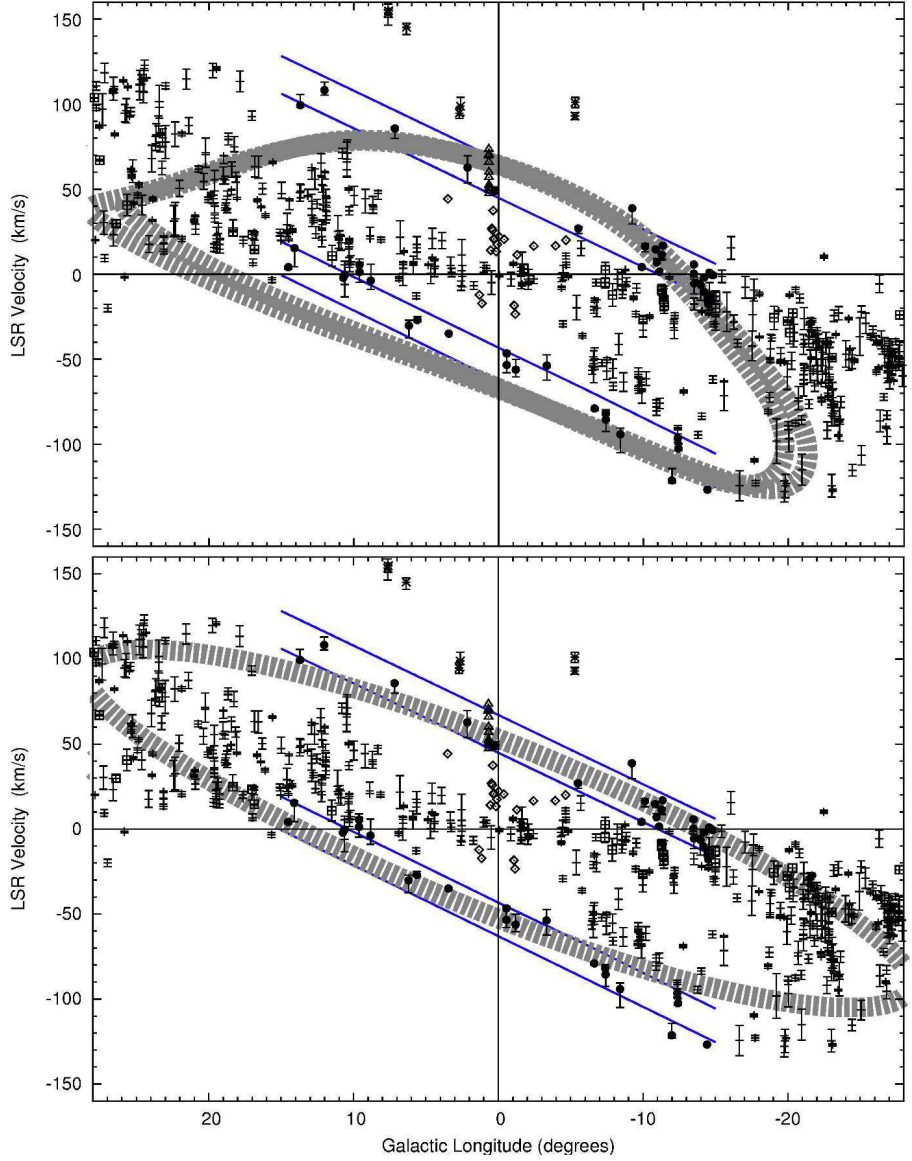} 
\caption{\small The elliptical streamlines corresponding to the 3-kpc arm (top) from \citet{peters75} and the expanding ring (bottom) from \citet{cohen76}. The top is based on two streamlines, one with a semi major axis of 4.9 kpc with an ellipticity of 0.69 and a fixed angular momentum of 770\,km\,s$^{-1}$\,kpc$^{-1}$, the other with a semi major axis of 4.6 kpc with an elliptiicity of 0.68 and a fixed angular momentum of 680\,km\,s$^{-1}$\,kpc$^{-1}$. The bottom is based on a radius of 4.0 kpc and a rotational velocity of 210\,km\,s$^{-1}$ with an expanding velocity of 53\,km\,s$^{-1}$.}
\label{petersellipse}
\end{center}
\end{figure}

The two original interpretations for the 3--kpc arm structure were that of an expanding ring \citep{rougoor64,kruit71, cohen76} or a non-expanding oval shaped resonance feature or elliptical streamlines \citep{devaucouleurs70, peters75}. Simple examples of these two structures are shown in relation to the Galactic bars in Figure \ref{barschematic} and loci are shown in the {\it l,v} domain for an example of each from the literature in Figure \ref{petersellipse}.
Development of both these interpretations was hindered at the time by the lack of a far-side counterpart and the inability to clearly trace structure at velocities less than the terminal. The CO detection of \citet{dame08} and the maser distribution allow us to re-examine these structures. The maser distribution of the MMB again suggests an oval structure in the {\it l,v} domain \citep{green10mmb2}. Although readily accounted for by a ring with an apparent expanding velocity, modelling and theoretical interpretation favour elliptical orbits/streamlines. 
Here we investigate this by firstly fitting the 3--kpc arm maser distribution with an elliptical model with constant angular momentum and, secondly, a circular ring model incorporating a component of radial velocity outwards from the Galactic centre. We fit only to the 3--kpc arm maser population as the individual sources towards the tangent points are dependent on the spiral arm model and rotation curve (as discussed in the Appendix). 

\begin{figure}
 \begin{center}
 \renewcommand{\baselinestretch}{1.1}
\includegraphics[width=14.5cm]{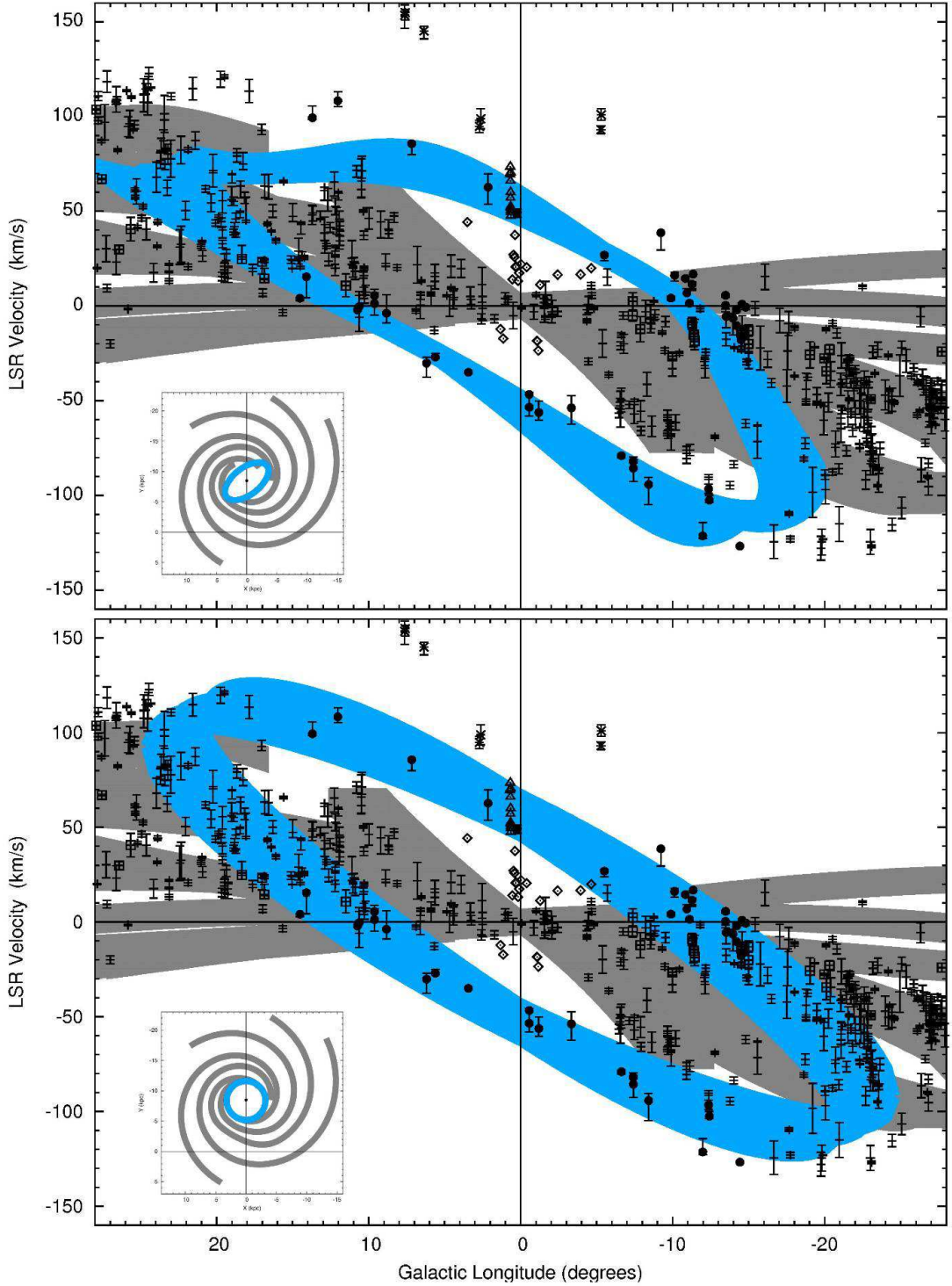} 
\caption{\small Longitude-velocity plot with a 3 kpc ellipse and ring. Symbols are the same as Figure \ref{lvdistribution}. The ellipse is 0.5 kpc thick in the radial direction and has a semi-major axis of 4.1 kpc, semi minor axis of 2.2 kpc, orientation of 38$^{\circ}$ and an angular momentum of 320\,km\,s$^{-1}$\,kpc$^{-1}$. The ring is 0.5 kpc thick in the radial direction and has an outward Galactocentric radial velocity between 40\,km\,s$^{-1}$ and 60\,km\,s$^{-1}$.}
\label{lvdistributionring}
\end{center}
\end{figure}

We first consider the case of an ellipse, with a ratio of semi--minor to semi--major axes of 0.50--0.80 \citep{peters75,sevenster99},  an orientation between 25$^{\circ}$ and 50$^{\circ}$ and a constant angular momentum of between 300 and 700\,km\,s$^{-1}$ kpc$^{-1}$. We assume the tangential velocity at any given point on the ellipse is given by the angular momentum divided by the Galactocentric radius of that point. The best fit is found for a semi--major axis of 4.1 kpc, an axis ratio of 0.54 (a semi--minor axis of 2.2\,kpc), an orientation of 38$^{\circ}$ and a rotational velocity of 320\,km\,s$^{-1}$ (a mean circular velocity of 106\,km\,s$^{-1}$). This accounts for 96\% of the 3--kpc arm sources and is shown (with a radial thickness of 0.5 kpc) overlaid on the maser distribution in Figure\,\ref{lvdistributionring}. The ellipse can approximate the parallel sections of the 3--kpc arms well, but does not associate the masers at positive longitudes ($>$15$^{\circ}$) with high velocities.
 
\begin{figure}
 \begin{center}
 \renewcommand{\baselinestretch}{1.1}
\includegraphics[width=14cm]{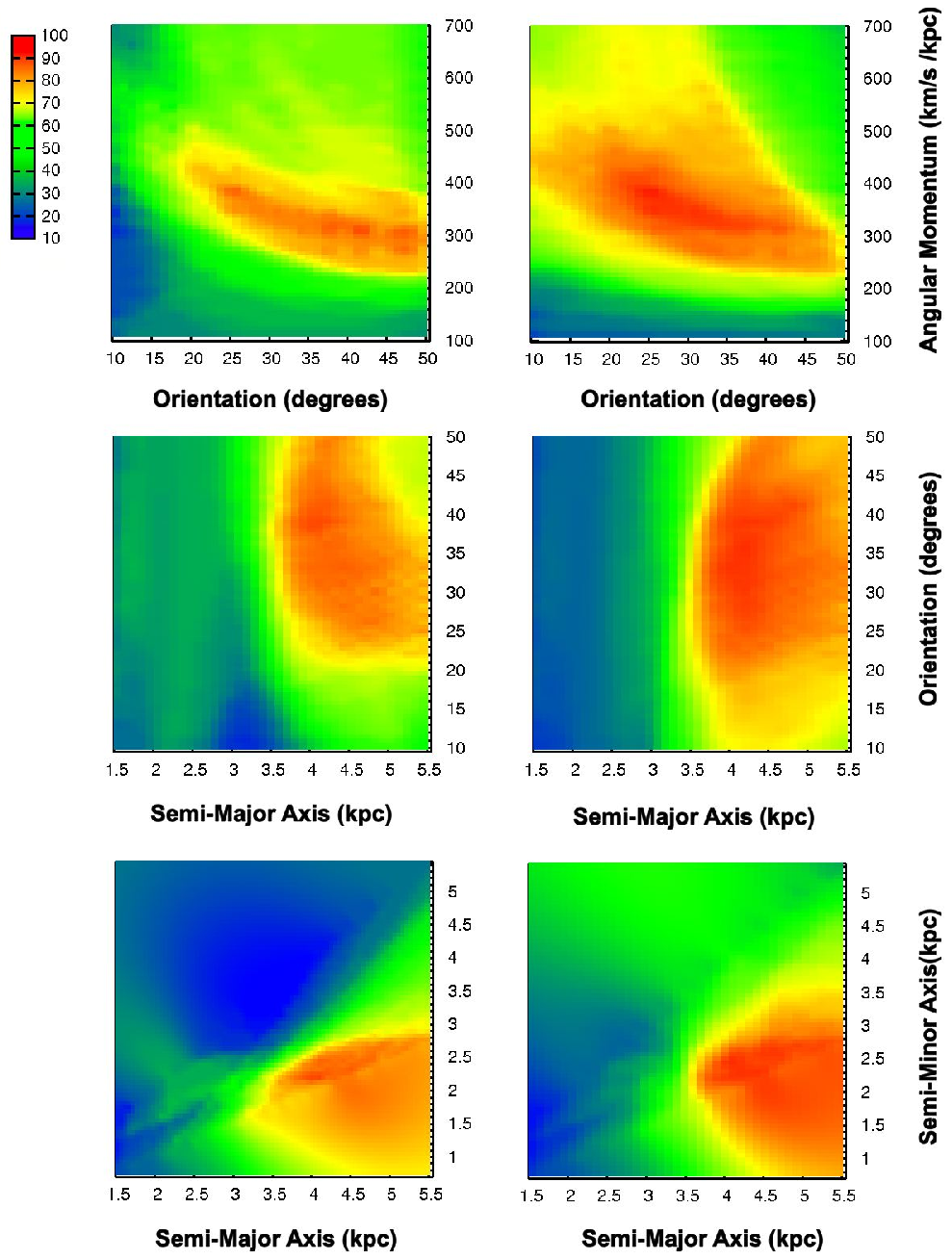}
\caption{\small Fraction of maser sources consistent with the 3--kpc ring model for different parameters of the ellipse. Colour scale represents 10 to 100\% source association. Left figures are for the association of 3--kpc arm sources only, right figures are for association with both the 3--kpc arm sources and the unassociated sources described in the text (with equal weighting to both). The top figures are for fixed semi major and semi minor axes, 4.1 kpc and 2.214 kpc respectively. The middle figures are for fixed semi-minor axis of 2.214 kpc and fixed angular momentum of 320 km\,s$^{-1}$ kpc$^{-1}$. The bottom figures are for a fixed orientation of 38$^{\circ}$ and angular momentum of 320 km\,s$^{-1}$ kpc$^{-1}$.}
\label{paramvary2}
\end{center}
\end{figure}

The effect on the level of association with varying parameters of the elliptical ring model is shown in the left panels of Figure \ref{paramvary2}. In this Figure, each pixel within a panel represents an ellipse model for a given set of parameters (Semi--major and --minor axes, orientation and angular momentum) and the colour scale represents the fraction of masers associated with the model. Extending the allowed range of angular momenta does not improve the fit, with the best fits confined within a limited range of $\sim$100\,km\,s$^{-1}$ kpc$^{-1}$ about 300 to 400\,km\,s$^{-1}$ kpc$^{-1}$, dependent on the orientation of the ellipse. Extending the length range of the axes gives much poorer fits below 3.5 kpc, but has comparable association at longer radii (although as noted radii larger than 4.5 kpc are increasingly unrealistic as they result in tangential velocities exceeding the H{\sc i} terminal velocities). Varying the orientation of the ellipse below 20$^{\circ}$ produces poorer fits. The effect of changing the orientation and the angular momentum is somewhat linked in the {\it l,v} domain, with a smaller orientation but larger angular momentum producing similar results to a larger orientation but smaller angular momentum. Decreasing the ratio of semi--minor to semi--major axes below 0.5 (i.e. increasing the level of ellipticity) produces a much poorer fit and increasing the ratio of axes much beyond 0.8 (making the ring increasingly circular) introduces the need for the radial motion of the circular ring scenario. If the unassociated sources described at the start of Section \ref{lvintro} are included in the fitting, then the best fit to the combined source sample is for a semi--major axis of 4.2 kpc, an axis ratio of 0.53 (a semi--minor axis of 2.2\,kpc), an orientation of 28$^{\circ}$ and a rotational velocity of 370\,km\,s$^{-1}$ (a mean circular velocity of 121\,km\,s$^{-1}$). This accounts for 91\% of the sources. The variation in the fraction of sources associated with the ring models when including the unassociated sources is shown in the right panels of Figure \ref{paramvary2}. 

For the second case of the circular ring, we take radius, radial velocity from the Galactic centre, and rotational velocity as free parameters. We assume the ring will start near the end of the bar, so allow the radius to range between 2.5 and 4.5 kpc. We vary the Galactocentric radial velocity between 40 and 60\,km\,s$^{-1}$ in accord with the range of velocities seen in the CO emission of the 3--kpc arms \citep{dame08}. In accordance with the proper motion of 9.621+0.196 \citep{sanna09} and our expectation of a lower rotational velocity towards the inner Galaxy \citep[e.g.][]{caswell10mmb1}, we adopt a circular velocity between 40 and 60\,km\,s$^{-1}$ lower than the Galactic rotation (246\,km\,s$^{-1}$). The highest level of source association is found for a radius of 3.4 kpc, a Galactocentric radial velocity of 48\,km\,s$^{-1}$ and a circular rotational velocity of 201\,km\,s$^{-1}$. This also accounts for 96\% of the 3--kpc arm sources and is shown (with a radial thickness of 0.5 kpc) overlaid on the maser distribution in Figure\,\ref{lvdistributionring}. 

\begin{figure}
 \begin{center}
 \renewcommand{\baselinestretch}{1.1}
\includegraphics[width=14cm]{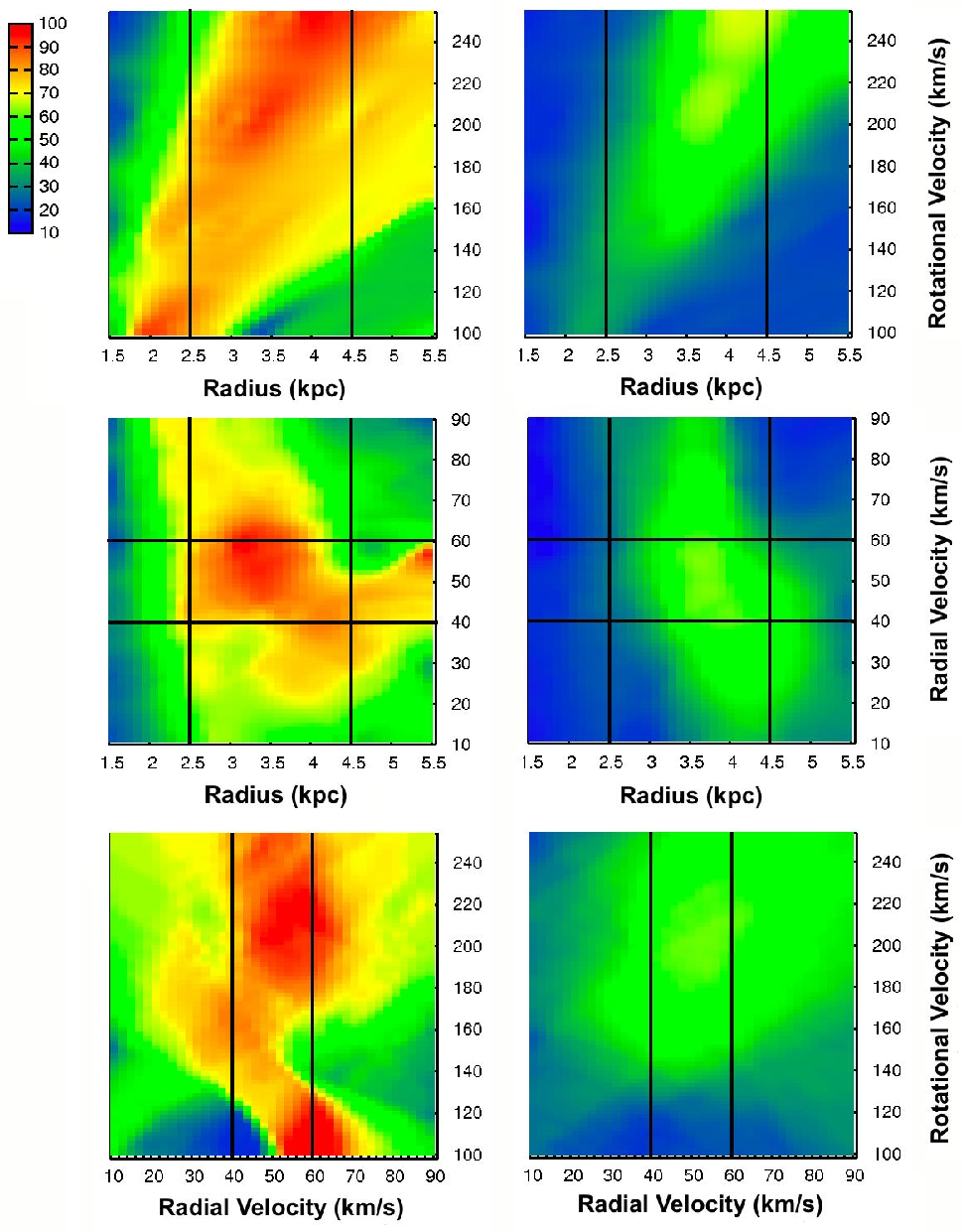}
\caption{\small Fraction of maser sources consistent with the 3--kpc ring model for different parameters of the ring. Colour scale represents 10 to 100\% source association. Left figures are for the association of 3--kpc arm sources only, right figures are for association with both the 3--kpc arm sources and the unassociated sources described in the text (with equal weighting to both). The top figures are for a fixed outward Galactocentric radial velocity of 48\,km\,s$^{-1}$, the middle figures are for a fixed circular velocity of 201\,km\,s$^{-1}$ and the bottom figures are for a fixed radius of 3.4 kpc. Black lines delineate likely boundaries for parameters from models and observations (see text for details). }
\label{paramvary}
\end{center}
\end{figure}

The effect on the level of association with varying parameters of the circular ring model is shown in the left panels of Figure \ref{paramvary}. As per Figure \ref{paramvary2}, each pixel within a panel represents a ring model for a given set of parameters (Galactocentric radius of the ring, outward Galactocentric radial velocity and circular velocity) and the colour scale represents the fraction of masers associated with the model. Extending the allowed range of radial velocities does not alter the best fit. Extending the range of radii gives a better fit for a larger radius (5.5 kpc) but this is too long physically (HI terminal velocities limit the possible length to $\sim$4.5 kpc). Varying the circular rotational velocity by a wider range gives marginally better fits (a few \%) for higher circular velocities at a given radius. Although not formally part of the fit, the ring model also accounts for the tangent sources (Section \ref{3kpctan}), the extreme negative sources between -15 and -24$^{\circ}$ longitude and the extreme positive sources between longitudes +15 and +24$^{\circ}$. If the unassociated sources described at the start of Section \ref{lvintro} are included in the fitting, then the best fit to the combined source sample is for a radius of 3.8 kpc, a radial velocity from the Galactic centre of 47\,km\,s$^{-1}$ and a circular velocity of 204\,km\,s$^{-1}$.  However this is a poorer fit to the 3--kpc arm sources between $\pm$15$^{\circ}$. The variation in the fraction of sources associated with the ring models when including the unassociated sources is shown in the right panels of Figure \ref{paramvary}. 

\subsubsection{Previous models of the structure}
Numerous alternative structure models have been previously explored. Through an N-body model and comparison with OH/IR stars, \citet{sevenster99} proposed an alternative elliptical ring. Although this correlated with the near 3--kpc arm known at the time, it is not a good match to the far 3--kpc arm or the $-$22$^{\circ}$ tangent. This model, which has a constant circular velocity rather than constant angular momentum struggles to produce the parallel nature of the near and far 3-kpc arms, requiring much longer radii than are reasonable to fit the distribution. If, instead of a continuous structure, the 3--kpc arms are considered to be two separate `lateral' arms, can the maser population be accounted for? Several authors have explored two lateral arms, recent examples being \citet{bissantz03} and \citet{rodriguez08} (the first a smoothed particle hydrodynamics model based on near infrared luminosities of the Diffuse Infrared Background Experiment (DIRBE)  in comparison with the CO {\it l,v} distribution and the second a sticky particles model based on star counts from the Two Micron All Sky Survey (2MASS)), but neither of these models account for more than about 30\% of the 3--kpc arm maser population (a third of the level of association of either the expanding ring or the ellipse). More complex asymmetric models which incorporate several arm-like features emanating from the bar fare no better: the models of  \citet{mulder86} (a two dimensional quasi-steady state model with near and far inner Lindblad radius arms and inner ultra-harmonic arms),  \citet{englmaier99} (a quasi-equilibrium flow model based on COBE/DIRBE near infrared luminosities) and  \citet{fux99} (a three dimensional N-body model based on COBE/DIRBE near infrared luminosities incorporating the 135\,km\,s$^{-1}$ arm and connecting arms), each associate no more than 30\% of the 3--kpc arm masers.

\begin{figure}
 \begin{center}
 \renewcommand{\baselinestretch}{1.1}
\includegraphics[width=10cm]{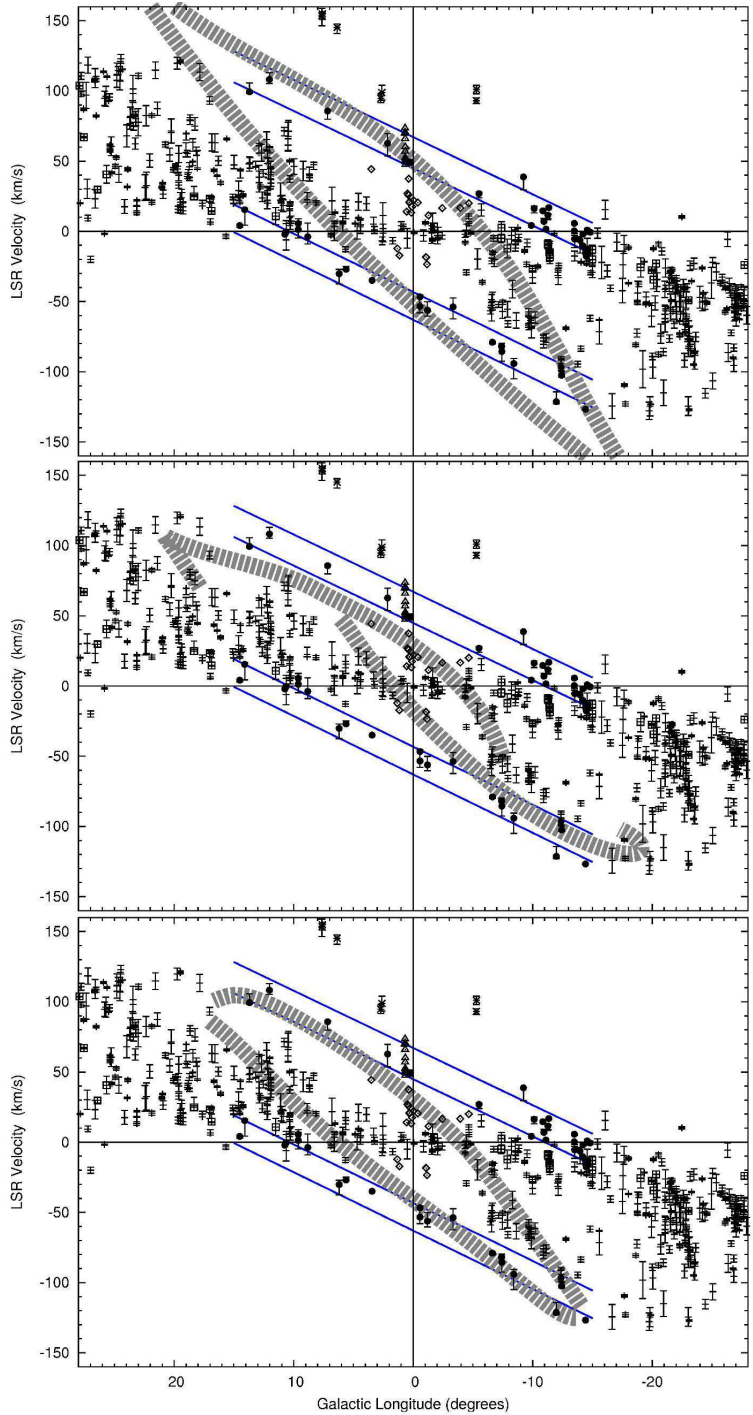}
\caption{\small Comparison of the 6.7--GHz methanol masers with model inner Galaxy structures. Symbols are the same as Figure \ref{lvdistribution}.  The blue lines delineate the region identified in CO emission as the 3--kpc arms by \citet{dame08}. TOP: \citet{sevenster99} elliptical ring with a bar orientation of 44$^{\circ}$ and a corotation radius of 4.5\,kpc. MIDDLE: \citet{bissantz03} combination of elliptically orbiting gas with a bar orientation of 20$^{\circ}$ and a corotation radius of 3.4\,kpc.
BOTTOM: \citet{rodriguez08} lateral arms with a combination bar of orientations of 20$^{\circ}$ and 75$^{\circ}$ and a corotation radius of 5--7\,kpc.}
\label{modelcomparisonpart1}
\end{center}
\end{figure}

\begin{figure}
 \begin{center}
 \renewcommand{\baselinestretch}{1.1}
\includegraphics[width=10cm]{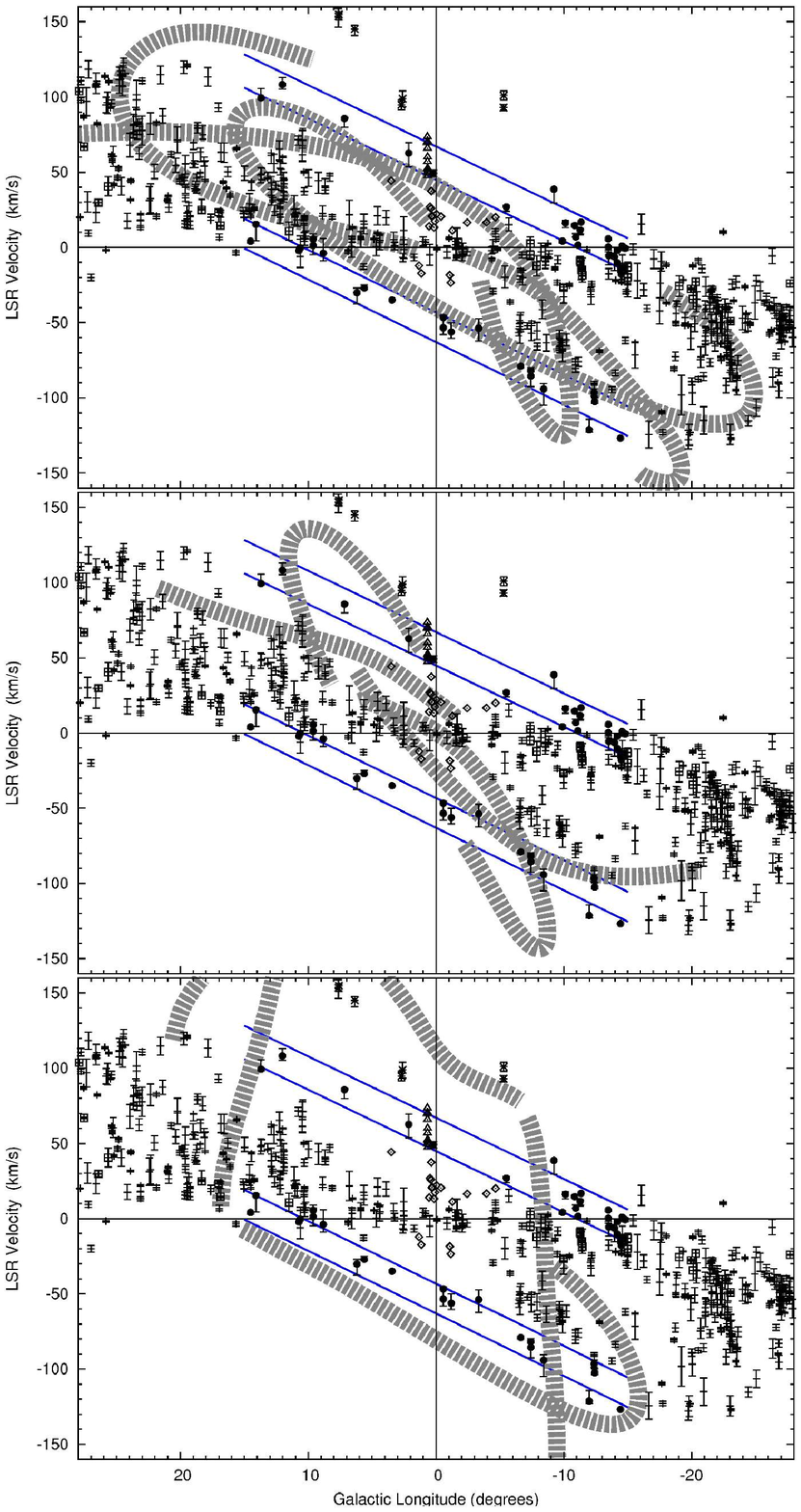}
\caption{\small Comparison of the 6.7--GHz methanol masers with model inner Galaxy structures continued. Symbols are the same as Figure \ref{lvdistribution}.  The blue lines delineate the region identified in CO emission as the 3--kpc arms by \citet{dame08}.
TOP: \citet{mulder86} four inner arm-like features: near and far inner lindblad radius arms and inner ultra-harmonic arms. This model has a 20$^{\circ}$ bar orientation and corotation radius of 4\,kpc.
MIDDLE: \citet{englmaier99} four inner arm-like features from the bar ends with a bar orientation of 20$^{\circ}$ and a corotation radius of 3.4 kpc. BOTTOM: \citet{fux99} four inner arm-like features: the 3--kpc arm, the 135\,km\,s$^{-1}$ arm and the near and far connecting arms. This model has a bar orientation of 25$^{\circ}$ and a corotation radius of 4$-$4.5\,kpc.}
\label{modelcomparisonpart2}
\end{center}
\end{figure}

The details of all the models discussed above are summarised in Table \ref{modeltable} and their loci are shown in Figures \ref{modelcomparisonpart1} and \ref{modelcomparisonpart2}. The models presented are as published, incorporating the authors' chosen best parameters (e.g. pattern speeds, bar orientations, solar distances) and also including a 10\,km\,s$^{-1}$ error margin in the loci of the structures. A direct comparison between the types of models is complicated by the simple fact that the four arm models cover more of the {\it l,v} domain, increasing their likelihood of associating sources. However, within the elliptical/lateral two arm models, that of \citet{rodriguez08} provides the best fit to the sources. This model has a large corotation radius of 5$-$7 kpc, as a result of a slow bar (where that bar is made up of both a bulge and longer component). The inclusion of a long bar component, clearly influential on high-mass star formation (Section \ref{barimp}), may be responsible for this higher association. Within the more complex multi-arm models, the resonance features of \citet{mulder86} provide the highest level of association. The lack of significant 6.7--GHz methanol maser emission towards the 135\,km\,s$^{-1}$ and `connecting' arms in these models suggests these may be primarily gas features, not undergoing high-mass star formation. This may be due to the broad range of velocities, which both features exhibit, indicating highly turbulent material. We similarly do not see 6.7--GHz methanol maser emission towards the broad range of velocities of material towards the Galactic Centre (the methanol velocity range is $\sim$200\,km\,s$^{-1}$ compared with the CO and H{\sc i} spanning $\sim$500\,km\,s$^{-1}$) and no emission is seen towards Bania's `Clump 2' \citep{bania77}, a feature with a high range of velocities, but it is seen towards the smaller velocity range of Bania's `Clump 1' \citep{green10mmb2}.

\subsubsection{Summary of the 3--kpc arm structure} 
In addition to the near and far 3--kpc arm population we find the maser {\it l,v} distribution demonstrates a prominent feature near $-$22$^{\circ}$ longitude which we interpret as the approximate tangent of the 3--kpc arms. Given the association of 6.7-GHz methanol masers with the parallel sections of the 3--kpc arms and the tangent, we can expect the entire structure to exhibit methanol maser emission and form an approximately oval structure in the {\it l,v} domain. When this structure is combined with the spiral arms and other features, essentially all the maser emission is accounted for. This oval structure can either relate physically to an elliptical ring or a circular ring with radial motion (`expansion'). The average surface density of sources in the Galactic plane, assuming an annulus of 0.5 kpc centred on 3.5 kpc, is 6 and 7 sources per kpc${^2}$ for the near and far portions of the arms ($\pm$15$^{\circ}$ longitude) respectively. This is comparable to the expected `smeared' density of sources in the spiral arms \citep{caswell10mmb1}. The average density along the tangent (excluding sources associated with other structural features, i.e. limiting the velocity range to $-$60\,km\,s$^{-1}$ to $-$85\,km\,s$^{-1}$) is 22 sources per kpc${^2}$. This higher density can be accounted for by the presence of a spiral arm origin, which we examine in Section \ref{armorigins}. 

The existing models of the dynamics on the inner Galaxy account for $\le$34\% of the 3--kpc arm sources. In comparison, a simple elliptical ring or `expanding' circular ring  account for of the order of 95\%. Although there are several models for an `expanding' circular ring close to the centre of our Galaxy (within a Galactocentric radius of $<$1 kpc) \citep[e.g.][]{scoville72, kaifu72, kaifu74, bally87}, and numerous examples of barred external galaxies with rings \citep[e.g.][]{buta86, buta96,treuthardt09}, dynamical models of the Milky Way since the early examples in the 1970s do not favour such a feature. Some radial motion may be a result of the influence of the radius of corotation resonance: a product of the transition between the elliptical orbits which follow the bar inside the radius of corotation resonance and the circular orbits which exist outside, with the 3--kpc ring the interface region between the two, pulling elliptically orbiting material into a circular orbit (an early example is that of \citealt{shane72}). The more favoured picture (in terms of dynamical modelling) of the elliptical ring or stream lines can account for the parallel sections and to some degree the negative tangent, but cannot readily account for the positive longitude masers with large velocities. The 6.7--GHz masers, as tracers of the structure, have the potential to fully define the spatial and kinematic structure of the 3--kpc ring, since further astrometric measurements  will allow parallax distances (thus a direct measure of the ellipticity of the ring) and the full three dimensional velocity behaviour.

The spiral arms are believed to originate at approximately the radius of corotation resonance of the Galactic bar \citep[e.g.][]{lopez99,englmaier99}, where the Galactic material is rotating at the same speed as the pattern speed of the bar, and radial oscillation is zero  \citep{lindblad74}. Highly elliptical orbits exist within the radius of corotation resonance, in contrast to circular orbits outside \citep{englmaier06}. This requires that the 3--kpc arms should be either a more elliptical structure interior to the radius of corotation resonance, with orbits following the bar  \citep[e.g.][]{contopoulos80, englmaier00} or a more circular structure close to the radius of corotation resonance, similar to resonance rings seen in external galaxies, such as NGC2523 and NGC4245 \citep[e.g.][]{buta99,treuthardt09}. As discussed in Sections \ref{3kpctan} and \ref{3kpcstrutfit}, the maser distribution leads us to believe the long thin bar is the most influential on star formation, with a bar semi-major axis of $\sim$3.4 kpc and orientation of $\sim$45$^{\circ}$, and the 3 kpc arms appear to be well approximated by an elliptical ring (or circular with radial velocity components resulting from the resonance). This indicates a radius of corotation resonance of $\sim$4.0 kpc, and a possible location for the origins of the spiral arms.

\begin{table} 
\center
\caption{Summary of comparison of maser distribution with models of the inner Galaxy. Details are given for the six models together with percentage association with the 3 kpc arm maser distribution. CR is the corotation resonance. References are: M86 = \citet{mulder86}; E99 = \citet{englmaier99}; F99 = \citet{fux99}; S99 = \citet{sevenster99}; B03 = \citet{bissantz03}; R08 = \citet{rodriguez08}.} 
\begin{tabular}{llcccc}
\hline
\multicolumn{1}{l}{Model} &  \multicolumn{1}{l}{Ref.} & \multicolumn{2}{c}{Galactic Bar} & \multicolumn{1}{c}{Radius} & \multicolumn{1}{c}{Level of}\\
\multicolumn{1}{l}{Description} & & \multicolumn{1}{c}{orientation} & \multicolumn{1}{c}{semi-major axis}& \multicolumn{1}{c}{of CR} & \multicolumn{1}{c}{Association}\\
&&( $^{\circ}$)&(kpc)&(kpc)&(\%)\\
\hline
2 lateral arms &R08& 20 \& 75 & 1.0 \& 0.2 &5.0$-$7.0&33\\
4 inner arms &M86   & 20 & 4.0 &8.4 &31\\
elliptical ring &S99  & 44 & 2.5 &4.5 &25\\
4 inner arms   &E99& 20 & 3.0 & 3.4 &20\\
2 lateral arms &B03& 20 & 1.8 &3.4 &16\\
4 inner arms  &F99 & 25 & 3.5 &4.0$-$4.5&9\\
\hline
\end{tabular} 
\label{modeltable}
\end{table}

\subsection{The spiral arm origins}\label{armorigins} 
Although the tangents of all the spiral arms have been recognised in CO, H{\sc i} and other tracers such as CS \citep[e.g.][]{bronfman92} and IR star counts \citep[e.g.][]{benjamin05}, identification of the arm origins, which fall within the terminal velocity envelope rather than at its boundary where the tangent points lie, has been elusive. This has meant models generally incorporate arbitrary or inferred arm origins within the {\it l,v} domain: the two `major' arms (Perseus and Crux-Scutum) originate at opposite ends of the Galactic bar and the two `minor' arms (Norma and Carina-Sagittarius) are anywhere between a few degrees and 90$^{\circ}$ offset, either starting from the radius of corotation resonance directly or branching from the two main arms. However, studies of other Galaxies and dynamical models show that the major arms do not necessarily start directly at the bar end, but can be seen offset by several degrees of Galactic azimuth \citep[e.g.][]{sandage94,russeil03}.

\subsubsection{Spiral arm induced density enhancements} 
As a product of interstellar material undergoing the influence of density waves, spiral arms are regions of dense material and star formation. External galaxies show the spiral arms start close to the end of the bar or a ring surrounding a bar and, as already mentioned, dynamical models place the arms starting at the radius of corotation resonance ($\sim$ 4 kpc). These regions are likely to experience significant compression of material such that star formation, induced by density changes in the ISM, is likely to be enhanced. The remaining high density regions of masers within $\pm$28$^{\circ}$ may thus be explained by the origins of the spiral arms. The exception is between longitudes +6$^{\circ}$ to $-$6$^{\circ}$ where all the spiral arms pass and there are sources on the solar circle (zero line-of-sight velocities) together with sources associated with the Galactic Centre Zone. In this region an overabundance is to be expected even in the absence of spiral arm origins.

High positive velocities at longitudes 25$^{\circ}$ to 35$^{\circ}$ have been considered the approximate start of the Crux-Scutum arm, with a region of massive star formation inferred from a sequence of IR observations \citep{hammersley94,garzon97,lopez99,hammersley00,cabrera08} and a massive young cluster of red supergiants \citep{davies07,clark09,negueruela10}. We see a high density of methanol masers in this region (1 bin in excess of 5\,$\sigma_{poisson}$ in Figure\,\ref{lvdensity}). The Crux-Scutum arm is the only arm with velocities above 60\,km\,s$^{-1}$ at these longitudes, it is known to have a tangent at $\sim$35$^{\circ}$, and originate at longitudes smaller than this. The lower estimate of the radius of corotation resonance is 3.5 kpc, which is tangential at 24$^{\circ}$. Hence the arm theoretically originates between 24$^{\circ}$ and 35$^{\circ}$ and we interpret the dense cluster of masers at $\sim$26$^{\circ}$ as observational evidence of the start of this arm.  Assuming the longitudes 24$^{\circ}$ and 35$^{\circ}$ mark the boundary of this region, and that it is a portion of an annulus with a thickness of 1 kpc centred at the corotation resonance radius of 4 kpc, the origin has a density of 11 sources per kpc$^{2}$. The discovery of the massive young red supergiant cluster in this region \citep{figer06} implied substantial starburst activity 10--20 Myr ago \citep{davies07}, but the significant presence of 6.7-GHz methanol masers indicates there is still substantial ongoing star formation.

As mentioned in Section \ref{3kpctan}, the methanol maser distribution demonstrates a dense ridge of sources near $l = $ $-$22$^{\circ}$ which is coincident with the negative longitude tangent of a $\sim$3.5 kpc ring. The density of masers within this region is 22 sources per kpc$^{2}$ (Section \ref{3kpcstrutfit}) compared to the average of both the 3 kpc ring and spiral arms which is 6 or 7 sources per kpc$^{2}$. This leaves 15 or 16 sources per kpc$^{2}$ which could be accounted for with the origin of a spiral arm. The Perseus arm is the only spiral arm in conventional models extendable to this region and is likely to correspond to the most negative velocities of the ridge. 

We now turn to the minor arms, Carina-Sagittarius and Norma. From fitting a logarithmic spiral to objects associated with the Carina-Sagittarius arm, and extrapolating towards the Galactic centre, the origin of this arm is often located within the fourth Galactic quadrant within about 10$^{\circ}$ of the Galactic centre \citep[e.g.][]{russeil03}. We observe a high density clump near longitude of $-$8$^{\circ}$ (1 bin above 5\,$\sigma_{poisson}$ in Figure\,\ref{lvdensity}). This is either a very dense and active region of high mass star formation within the Norma arm (at non-tangential longitudes and more negative velocities than expected) or it is more likely the indication of the start of the Carina-Sagittarius arm and the overlapping of the {\it l,v} parameters of this arm with those of the Norma arm (with the further implication that the Norma arm branches from the Perseus arm). The starting point of the Norma arm is less clear. It is often speculated to originate in the first Galactic quadrant, approximately opposite to the other minor spiral arm Carina-Sagittarius \citep[e.g.][]{ortiz93}. We see a dense cluster of sources from longitudes +10$^{\circ}$ to +12$^{\circ}$ extending to positive velocities beyond the Carina-Sagittarius and Crux-Scutum arms, which could be interpreted as the start of the Norma arm. If the Norma and Carina-Sagittarius arms branch from the major arms outside the radius of corotation resonance, rather than starting near the ring at the radius of corotation resonance, the conditions of the ISM may be less compressed and the level of star formation will not be as high as the origins of the major arms (the branching scenario is a more logical scenario geometrically if the ring is elliptical).  

The associations of the density enhancements are summarised in Table \ref{resotable} together with an approximation of the density of sources per kpc$^{2}$. We identify the high density regions of masers not associated with the bar or the 3--kpc arms as enhanced star formation at the starting points of the spiral arms: Crux-Scutum at approximately +26$^{\circ}$ (velocities circa 100\,km\,s$^{-1}$); Norma at approximately +12$^{\circ}$ (velocities circa 50\,km\,s$^{-1}$); Carina-Sagittarius at approximately $-$8$^{\circ}$ (velocities circa $-$55\,km\,s$^{-1}$); and Perseus at approximately $-$22$^{\circ}$ (velocities circa $-$55\,km\,s$^{-1}$). We find the Crux-Scutum and Carina-Sagittarius arm origins to have densities of 10--15 sources per kpc$^{-2}$, similar to the densities seen at the ends of the Galactic bar. The densities of the Norma and Perseus arms are confused by additional arms lying on the line-of-sight and the 3--kpc arm tangent. The masers contained within the spiral arm origins have a median latitude of 0.0$\pm$0.1$^{\circ}$, are typically contained within a range of about 30\,km\,s$^{-1}$ and have median fluxes in the range 6 to 10 Jy.

\section{Conclusion}
 We have examined the distribution and density of 6.7--GHz methanol masers in the longitude-velocity domain. Both the longitude-velocity distribution and its structure function demonstrate the presence of structures on small ($<$0.03$^{\circ}$) and large ($>$3$^{\circ}$) scales. Through smoothing the density distribution on the small scales and binning on the large scales we identify statistically significant dense regions of masers indicative of enhanced high-mass star formation within Galactic scale structures. The maser distribution supports the presence of a long thin bar inclined at angle of $\sim$45$^{\circ}$ to the Sun-Galactic centre line of sight. The lack of methanol emission from a short, bulge associated bar implies the two Galactic bars represent different ages: the short boxy bulge bar an older structure and the long thin bar a younger structure, undergoing current high-mass star formation. We identify a prominent tangent of the 3--kpc arms near $-$22$^{\circ}$ and find the maser distribution of the 3--kpc arms is readily associated with a continuous ring structure in the longitude-velocity domain. High densities of masers identify the approximate starting points of the spiral arms: the major arms, Crux-Scutum at +26$^{\circ}$ and Perseus at -22$^{\circ}$, slightly offset from the bar ends; the minor arms, Norma at 12$^{\circ}$ and Carina-Sagittarius at -8$^{\circ}$, possibly branching from the major arms. The 3--kpc arm ring and spiral arm origins, combined with the spiral arms themselves, account for essentially all the 6.7--GHz methanol masers and their density enhancements within longitudes $\pm$28$^{\circ}$. 6.7--GHz methanol masers clearly delineate many of the important structures of our Galaxy and provide a new observational basis to constrain dynamical models. Multi-wavelength studies and astrometric distances through VLBI will further enhance the importance of this species of maser as a tool in understanding the structure of our Galaxy.
 
{\it Facilities:} \facility{Parkes (Methanol Multibeam)}.

\appendix

\section{Minor implications of uncertainties in spiral arm shape and rotation parameters}\label{Discussion} 
As noted in Section \ref{lvintro} we adopt the \citet{taylor93} spiral arms and the \citet{brand93} rotation curve, together with a Galactocentric solar distance (R$_{\odot}$) of 8.4 kpc and a circular rotation of the Sun ($\Theta_{\odot}$) of 246\,km\,s$^{-1}$. We now discuss the potential implications of uncertainties in our assumptions regarding the shape of the spiral arms and the rotational parameters of the Galaxy. 

\subsection{Choice of spiral arm shape}\label{armchoice}
Our knowledge of the Galactic location and spatial properties of the spiral arms has been limited by the accuracy of the distances of objects tracing the spiral arms, such as H{\sc ii} regions and Giant Molecular Clouds. The often cited model of \citet{georgelin76} is based on interpretation of star forming regions with assigned distances, many of which are kinematic distances, known to suffer significant error margins \citep[see for example][]{gomez06,reid09}. Models are often made to fit the observed tangential velocities of atomic (H{\sc i}) and molecular (CO) emission, interpolating the structure between these extremes to approximate what is seen observationally \citep[e.g.][]{baba10}. As such, widely used models of the spiral arms, such as \citet{taylor93}, have changed minimally in the past 44 years since \citet{georgelin76}, so suffer the same uncertainties and some caution should therefore be taken with the location of the spiral arm loci. Statistical arguments summarising the models suggest there is a favoured picture of a logarithmic four arm spiral with a pitch angle of $\sim$12$^{\circ}$  \citep{vallee95, vallee02, vallee05, vallee08a}. The recent studies of \citet{russeil03} and \citet{hou09}, with logarithmic and polynomial fits respectively, favour a four arm model and specifically one which has two `major' and two `minor' arms. \citet{levine06} found concurring evidence for a four arm structure in the H{\sc i} emission in the outer Galaxy. Additionally, although the Galaxy appears to have an overall grand structure, spurs, such as the local `Orion arm', exist. 

It is the combination of observations at multiple wavelengths and the `piggybacking' of distances through associations of these \citep[e.g.][]{urquhart10}, that is helping to improve our current understanding and models. Ultimately however, the thousands of stellar distances of the upcoming NASA Gaia mission together with astrometric parallaxes of masers through VLBI (identifying the optically obscured star forming regions) will have the potential to truly delineate the spiral arms. Unfortunately, both of these are several years away, making the loci of models our current best estimate: until then the likely overall structure, number of arms and continuity of the arm loci will remain helpful in recognising potential spiral arm association of maser sources. 

\subsection{Choice of rotation curve}\label{rotationchoice}
As the \citet{taylor93} arms are largely based on those of \citet{georgelin76} and these were based on kinematic distances using a rotation curve, it is pertinent to adopt a similar rotation curve to convert the arm shapes to the {\it l,v} domain. \citet{georgelin76} used a rotation curve with a $\Theta$$_{\odot}$ of $\sim$250\,km\,s$^{-1}$ and an R$_{\odot}$ of 10\,kpc. When scaled to 8.5\,kpc this equates to a $\Theta$$_{\odot}$ of $\sim$220\,km\,s$^{-1}$, for which the \citet{brand93} rotation curve is a good approximation. However, we also tested the association of masers with the rotation curves of \citet{clemens85}, \citet{fich89}, \citet{mcclure07} and \citet{reid09}. With 1\,kpc thick spiral arms and a 7\,km\,s$^{-1}$ velocity tolerance in association we find: \citet{clemens85} accounts for 77\% of the sources with the spiral arms; \citet{fich89} linear curve 78\%; \citet{fich89} power-law curve 79\%; \citet{mcclure07} 80\%; and the flat rotation of \citet{reid09} accounts for 76\% of the sources. Although individual examples of associated sources change between the rotation curves, the total remains in the range 76$-$80\%, mirroring the conclusions of the previous section and emphasising that the inferred total number of spiral arm sources is largely unaffected by rotation curve choice.

\subsubsection{Solar Parameters}
In addition to the choice of rotation curve, Galactic kinematics are dependent on the parameters of solar motion and the local standard of rest. For the solar distance, $R_{\odot}$, the consensus appears to be 8.4 kpc, based on orbits of S0-2 stars implying 8.4$\pm$0.4 kpc \citep{ghez08} and 8.33$\pm$0.35 kpc \citep{gillessen09}. This is consistent with the parallax observations of water masers in Sgr B2 indicating 7.8$\pm$0.8 kpc \citep{reid09b} and the parallaxes of methanol masers indicating 8.4$\pm$0.6 kpc \citep{reid09}. 

The IAU standards of solar motion are implicitly incorporated in our maser LSR velocities. However, recent maser astrometric observations have prompted revision. Two of the parameters, solar motion towards the Galactic centre, $U_{\odot}$, and solar motion towards the north Galactic pole, $W_{\odot}$, have remained largely unchanged (10.3\,km\,s$^{-1}$ and 7.7\,km\,s$^{-1}$ respectively), but the solar motion in the direction of Galactic rotation, $V_{\odot}$, has not.
Originally $V_{\odot}$ was chosen to be 15.3\,km\,s$^{-1}$ (the IAU standard), then it was revised to 5.25$\pm$0.62\,km\,s$^{-1}$ based on stellar kinematics of the Hipparcos catalogue \citep{dehnen98}. However the dataset was re-examined, resulting in further revision to 12.24$\pm$0.47\,km\,s$^{-1}$ \citep{schonrich10}. This higher value of $V_{\odot}$, returning to a value close to the original IAU standard, is in line with the suggestions of \citet{reid09} and \citet{mcmillan10} based on maser proper motions. The revision to a higher value may account for the apparent motion of star forming regions counter to Galactic rotation \citep{bobylev10}.

Galactic circular rotation of the Sun, $\Theta_{\odot}$ has also been subject to recent revision. The IAU standard of $\Theta_{\odot}$, 220\,km\,s$^{-1}$, has been revised to a higher value based on the results of proper motions deduced from maser astrometry observations. The revised values are 254$\pm$16\,km\,s$^{-1}$ as estimated by \citet{reid09}, or either  246$\pm$30\,km\,s$^{-1}$ or 244$\pm$13\,km\,s$^{-1}$ as estimated by \citet{bovy09}. 

Therefore the current best estimates of the parameters are: $U_{\odot}$ = 11.1\,km\,s$^{-1}$; $W_{\odot}$ = 7.25\,km\,s$^{-1}$; $V_{\odot}$ = 12.2\,km\,s$^{-1}$; $\Theta_{\odot}$ = 246\,km\,s$^{-1}$; and $R_{\odot}$ = 8.4 kpc. The most significant difference is the increase in $\Theta_{\odot}$. We adopt the $R_{\odot}$ of 8.4 kpc throughout, but explored the impact of the other variations on our conclusions, and the effect is minimal. The adjustment of $\Theta_{\odot}$ has a reasonable impact on kinematic distance estimates, but as we are working in the {\it l,v} domain, it does not influence our maser parameters. It does affect the location of the spiral arm loci if we transfer them with this value rather than the IAU standard, but this only serves to shift the loci in velocity by $\sim$5-10\,km\,s$^{-1}$.

\bibliographystyle{mn2e} \bibliography{UberRef}

\clearpage

\end{document}